\documentclass[runningheads]{llncs}
\usepackage{graphicx}
\usepackage{amsmath}

\emergencystretch=\maxdimen
\usepackage{colortbl}

\begin{document}
\title{Policy-Based Reinforcement Learning for Assortative Matching in Human Behavior Modeling}
% If the paper title is too long for the running head, you can set an abbreviated paper title here 
%
\author{Ou Deng\inst{1}\orcidID{0000-0003-1172-1789} \and
Qun Jin\inst{1}\orcidID{0000-0002-1325-4275}}
\authorrunning{Ou Deng et al.}
\institute{Graduate School of Human Sciences, Waseda University, Japan\\
\email{dengou@toki.waseda.jp}\\
\email{jin@waseda.jp}\\
%\url{https://nislab.human.waseda.ac.jp/}
}
\maketitle              % typeset the header of the contribution
\begin{abstract}
This paper explores human behavior in virtual networked communities, specifically individuals or groups' potential and expressive capacity to respond to internal and external stimuli, with assortative matching as a typical example. A modeling approach based on Multi-Agent Reinforcement Learning (MARL) is proposed, adding a multi-head attention function to the A3C algorithm to enhance learning effectiveness. This approach simulates human behavior in certain scenarios through various environmental parameter settings and agent action strategies.
In our experiment, reinforcement learning is employed to serve specific agents that learn from environment status and competitor behaviors, optimizing strategies to achieve better results. The simulation includes individual and group levels, displaying possible paths to forming competitive advantages. This modeling approach provides a means for further analysis of the evolutionary dynamics of human behavior, communities, and organizations in various socioeconomic issues.

\end{abstract}
\section{Introduction}

The exploration of human behavior patterns has mainly relied on quantitative and qualitative actual investigation and analysis and social experiments under certain conditions. These traditional methods cannot avoid many practical limitations, such as the budgetary cost of conducting social surveys and experiments, the difficulty of obtaining data resources, and even ethical and moral constraints. Due to these limitations and difficulties, multiagent system (MAS) simulation and artificial intelligence methods are utilized to assist traditional social investigation and analysis. Even for some social experiments that cannot carry out by ethical or moral restrictions, a well-designed simulation can provide a certain degree of reliable results or help to do counterfactuals in social experiments.

Considering the high complexity of human behavior patterns, we choose a simple but fundamental human behavior -- assortative matching -- for our social experiments. The matching theory has essential applications in social and economic fields, e.g., labor market, industry planning, and international trade. 
This study was inspired by a coupling game conceived by Dan Ariely \cite{Ariely_upside_of_irrationality}, who claimed and guessed that the result would be an \textit{\textbf{equal matching}} when single men and women play such a coupling game under specific rules, without experimenting.

We design and implement simulation experiments with multi-agent systems and machine learning methods, which verify the common-sense-based \textit{equal matching}. In the designed virtual environment, our policy-based reinforcement method improves strategies to enable some agents to obtain competitive advantages over others. The experiments examine the transformation conditions of these competitive advantages and the transformation process, which are essential to the evolving dynamic of social communities.

\section{Related Works}

Policy-based reinforcement learning is a subfield of reinforcement learning focused on developing algorithms for learning control policies to manage agents in complex environments. The application of policy-based reinforcement learning to assortative matching in human behavior modeling is a relatively new research area.

This area of research emerged in the 2010s, when computer scientists and computational social scientists explored the use of reinforcement learning techniques to model and predict relationship formation in large-scale social networks. Early work focused on developing algorithms that can influence relationship formation by incentivizing individuals to form relationships with others who are similar to themselves in specific characteristics.

%"Playing Atari with Deep Reinforcement Learning" by
The use of deep neural networks in reinforcement learning was introduced by Mnih et al. from Google DeepMind \cite{DeepMind_2013}. In 2015, the same research group introduced the Deep Q-Network (DQN) algorithm \cite{DeepMind_2015}, which uses deep neural networks to approximate the action-value function. The algorithm demonstrated superhuman performance in many Atari games. In 2016, the Asynchronous Advantage Actor-Critic (A3C) algorithm \cite{DeepMind_2016_A3C} was introduced, which uses multiple parallel actors to interact with the environment and reduce the time needed for convergence.

Since then, the field has developed new policy-based reinforcement learning algorithms for assortative matching in online social networks, online dating platforms, and mobile communication networks. These algorithms have been applied to real-world data and provided new insights into the factors that shape relationships and social networks.

The Trust Region Policy Optimization (TRPO) algorithm was introduced by Schulman et al. of OpenAI \cite{TRPO}, which uses a trust region constraint to ensure stable updates to the policy. The Proximal Policy Optimization (PPO) algorithm \cite{PPO} was introduced in 2017, which combines ideas from policy gradient methods and trust region methods to provide a flexible and computationally efficient approach to policy optimization.

Recent research has explored the use of deep reinforcement learning techniques for policy-based assortative matching, which can improve the performance of algorithms and address challenges associated with traditional reinforcement learning algorithms. 

However, limited research applied policy-based reinforcement learning to assortative matching in human behavior modeling, a multidisciplinary field combining economics, sociology, and computer science techniques to study the factors that shape relationships and social networks. Most researches in this field have been based on analytical and computational models rather than reinforcement learning approaches.

\section{Methods}
\begin{figure}[ht]\centering 
\includegraphics[width=1.0\textwidth]{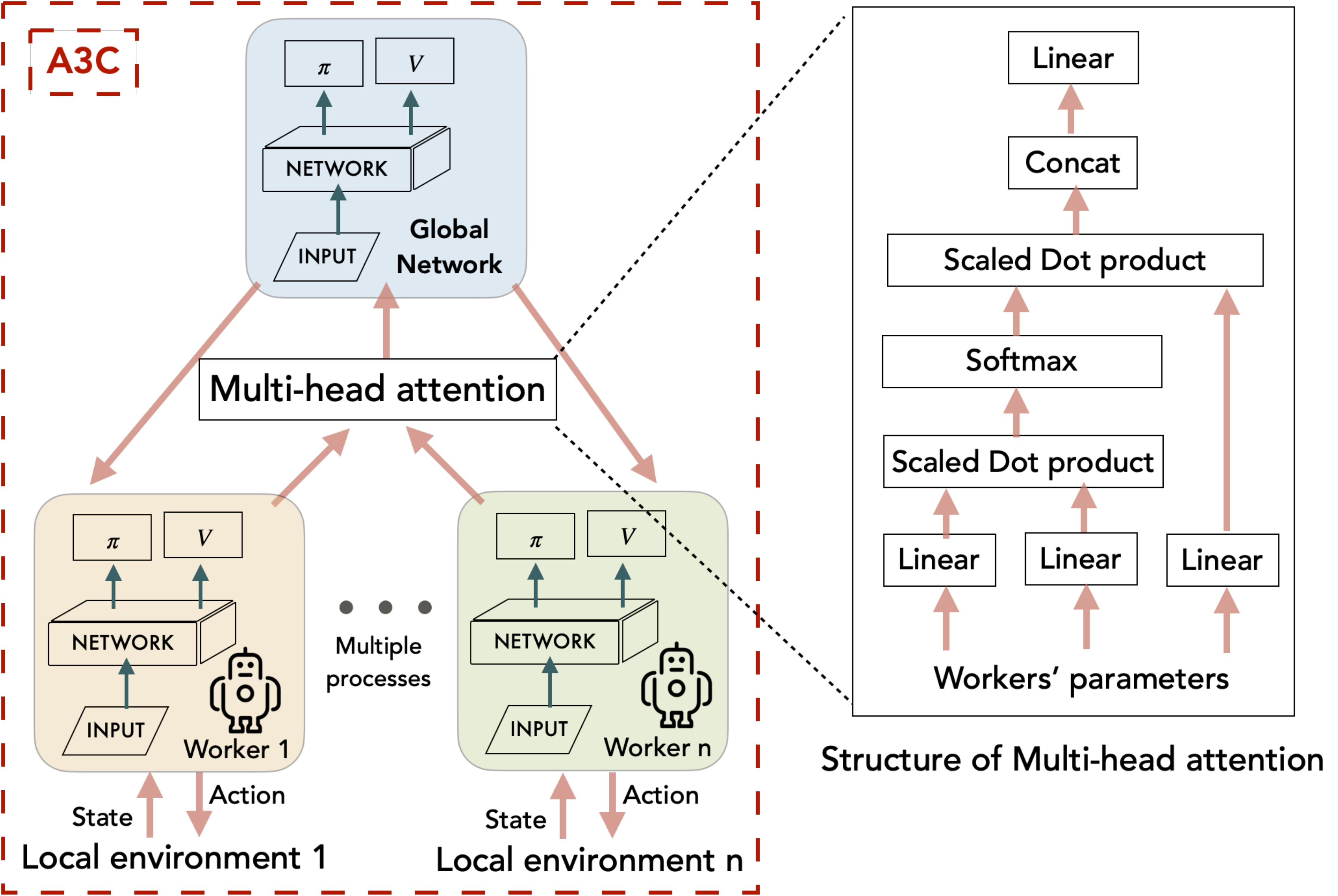}
\caption{Multi-Head Attention Asynchronous Advantage Actor-Critic (MA-A3C) network structure. The left section shows the general structure of MA-A3C. The right expand section shows the details of the multi-head attention mechanism.}
\label{fig_MA-A3C_1}
\end{figure}

This study presents a novel reinforcement learning algorithm, the Multi-Head Attention Asynchronous Advantage Actor-Critic (MA-A3C) algorithm, shown in Fig.\ref{fig_MA-A3C_1}. This approach integrates the Multi-Head Attention mechanism with the classic Asynchronous Advantage Actor-Critic (A3C) algorithm and is specifically designed to address assortative matching problems in reinforcement learning. Given the inherent complexities of human behavior patterns, the proposed MA-A3C algorithm provides a more effective solution for modeling these patterns. Additionally, the algorithm demonstrates scalability and has the potential for more comprehensive applications in research domains involving human behavior modeling.

The A3C algorithm is a widely recognized multi-agent reinforcement learning (MARL) technique that incorporates the actor-critic framework and parallel computing. This multi-agent model is particularly well suited for simulating human behavior, as it employs multiple parallel processing units to learn a policy asynchronously, thus enabling more efficient state space exploration.

Adding the MA-A3C algorithm can improve its performance. This is because the Multi-Head Attention function enables the algorithm to attend to different state space elements more flexibly and in a more fine-grained way. Multi-Head Attention is a common technique in deep learning that allows for creation of multiple attention mechanisms that can be applied to different input elements.

In reinforcement learning, Multi-Head Attention can be used to enable the A3C algorithm to attend to different states based on their relative importance. This can help improve the efficiency of the learning process by enabling the algorithm to focus on the most important elements of the state while ignoring the less important ones.

\subsection{Multi-Head Attention Function in A3C Algorithm (MA-A3C)}
Let $S$ be the state representation, $W_{i}$ be the weight matrix for the $i^{th}$ head, and $b_{i}$ be the bias vector for the $i^{th}$ head. The attention scores for the $i^{th}$ head can be calculated as the dot product of $S$ and $W_{i}$ with $\text{softmax}$ normalization: $a_{i} = \text{softmax}(S \cdot W_{i} + b_{i})$. Next, the attention scores are used to weight the features of the state representation: $S_{i} = a_{i} \cdot S$. Finally, the weighted features from each head are concatenated to produce the final enhanced state representation: $S_{\text{final}} = \left[S_{1}, S_{2}, ..., S_{H} \right]$, where $H$ is the number of heads. The enhanced state representation is then passed to the policy network to determine the action.

\subsection{General Procedures of MA-A3C Algorithm}

\subsubsection{(1) Enhancing state representation}
The state of the environment can be represented as a vector of features. This representation can be enhanced using normalization, dimensionality reduction, and feature selection techniques. Let $x_{t} \in R^{n}$ be the state of the environment at time $t$, and let $x'_{t} \in R^{m}$ be the enhanced representation, where $m$ is the number of features after enhancement. The enhancement process can be expressed as $x'_{t} = f(x_{t})$, where $f(\cdot)$ implements normalization, dimensionality reduction, and feature selection techniques.

For instance, normalization can be performed using the mean and standard deviation of the features. Dimensionality reduction can be achieved using principal component analysis (PCA) or singular value decomposition (SVD) techniques. Feature selection can be performed by selecting the most important features based on some criteria.

\subsubsection{(2) Calculating attention scores}
To compute attention scores for each head, we can use various methods such as dot product attention, scaled dot product attention, or MLP attention. Let $x'{t} \in R^{m}$ be the state representation and let $h$ be the number of heads. The attention scores for each head can be calculated as a set of scalar values $a_{t,1}, a_{t,2}, ..., a_{t,m}$, which indicate the relative importance of each feature in the state representation.

The dot product attention method can be used to calculate attention scores, where $a_{t,j} = (x'{t,j})^{T}\cdot w{j}$, and $x'_{t,j}$ is the $j^{th}$ feature of the state representation $x'_t$, and $w_j$ is the weight vector associated with the $j^{th}$ feature.

The attention scores are used to weight each head's state representation features. The weighted features can then be combined to form a context vector representing the state's multi-head attention representation. Alternatively, scaled dot product attention or MLP attention can be used for calculating the attention scores.

\subsubsection{(3) Normalizing attention scores}
In the A3C algorithm with multi-head attention function, attention scores are typically normalized using a softmax function to ensure that they sum to 1. This enables the attention mechanism to focus on the most significant parts of the state representation while ignoring the less important ones.

Let $a_t = [a_{t,1}, a_{t,2}, ..., a_{t,m}]$ be the vector of attention scores for time step $t$. The normalized attention scores can be calculated using the softmax function expressed in Eq.\ref{eq_normalizing}.
\begin{equation}
p_t = \text{softmax}(a_t) = \frac{\text{exp}(a_t)}{\sum_{j=1}^m \text{exp}(a_{t,j})}
\label{eq_normalizing} %Eq.\ref{eq_}
\end{equation}
The softmax function maps a vector of real numbers to a probability distribution, ensuring that the attention scores sum to 1, i.e., $\sum_{j=1}^m p_{t,j} = 1$.

Normalizing the attention scores allows the A3C algorithm to dynamically weight the features in the state representation based on their importance, resulting in a more effective representation of the state. The attention mechanism can focus on the most important parts of the state representation while ignoring the less important elements, which can improve the performance of the policy-based reinforcement learning algorithm.

\subsubsection{(4) Weighting features}
The attention scores are used to weight the features of the state representation. The weighted features are combined using a linear transformation to produce a new, enhanced state representation.

Let $s_t = [s_{t,1}, s_{t,2}, ..., s_{t,n}]$ be the state representation at time step $t$, and $p_t = [p_{t,1}, p_{t,2}, ..., p_{t,m}]$ be the corresponding normalized attention scores. The weighted state representation can be calculated by Eq.\ref{eq_weighting_features}.
\begin{equation}
z_t = \sum_{j=1}^m p_{t,j} \cdot s_{t,j}
\label{eq_weighting_features}
\end{equation}

This weighted state representation dynamically emphasizes the most important features of the state while ignoring the less important ones. The weighted state representation can then be fed into a neural network or machine learning model for further processing. This results in a more effective representation of the state that can be used for decision-making in the policy-based reinforcement learning algorithm.

\subsubsection{(5) Passing enhanced state to policy network}
The process of calculating attention scores, normalizing the scores, weighting the features, and passing the enhanced state to the policy network is repeated for each head. This allows the policy network to attend to different parts of the state space using multiple fine-grained attention mechanisms.

Let $\theta$ be the parameters of the policy network and $f_{\theta}(z_t)$ be the function representing the policy network that maps the enhanced state representation $z_t$ to a set of actions. The policy network can be formulated as Eq.\ref{eq_passing}.
\begin{equation}
a_t = f_{\theta}(z_t)
\label{eq_passing}
\end{equation}
where $a_t = [a_{t,1}, a_{t,2}, ..., a_{t,k}]$ is the vector of actions output by the policy network for time step $t$.

The policy network can be trained to maximize a reward signal that reflects the elements' performance in the environment.

The enhanced state representation is fed into the policy network, which outputs a set of actions that can be taken by the elements in the environment. This results in a more effective representation of the state and improved decision-making in the policy-based reinforcement learning algorithm.

\subsubsection{(6) Multi-head attention}
To enable the policy network to attend to different parts of the state space using multiple fine-grained attention mechanisms, we repeat the process of calculating attention scores, normalizing the scores, weighting the features, and passing the enhanced state to the policy network for each head.

Let $H$ be the number of heads in the multi-head attention function, and $h \in {1, 2, ..., H}$ be the index for each head. For each head $h$, we calculate the attention scores, $\alpha_{t,h} = [\alpha_{t,h,1}, \alpha_{t,h,2}, ..., \alpha_{t,h,n}]$, using a method such as dot product attention, scaled dot product attention, or multi-layer perceptron attention. We then normalize the attention scores using the softmax function to obtain the normalized attention scores by Eq.\ref{eq_MA}.
\begin{equation}
\beta_{t,h} = \text{softmax}(\alpha_{t,h})
\label{eq_MA} %Eq.\ref{eq_}
\end{equation}

We then weight the features of the state representation, $s_t = [s_{t,1}, s_{t,2}, ..., s_{t,n}]$, using the normalized attention scores to obtain the weighted features by Eq.\ref{eq_weighted_feature}.
\begin{equation}
z_{t,h} = \sum_{i=1}^{n} \beta_{t,h,i} s_{t,i}
\label{eq_weighted_feature} %Eq.\ref{eq_}
\end{equation}

Finally, we obtain the enhanced state representation for each head by concatenating the weighted features for each head. That is, $z_t = [z_{t,1}, z_{t,2}, ..., z_{t,H}]$, and this enhanced state representation can then be passed to the policy network to produce a set of actions, $a_t = f_{\theta}(z_t)$.

This process of applying the multi-head attention function in the A3C algorithm is repeated for each head to allow the policy network to attend to different parts of the state space using multiple fine-grained attention mechanisms.

\subsubsection{(7) Updating policy network}
The policy network is updated using the gradient of the policy objective, which is calculated using the observed rewards and estimated state values. The policy objective is defined as the difference between the expected return and the estimated value of each state, and the gradient of this objective is used to update the policy network.

Let $J(\theta)$ be the policy objective, where $\theta$ is the set of parameters of the policy network; $R_t$ is the reward observed at time step $t$; $V(s_t;\phi)$ is the estimated value of state $s_t$; $\gamma$ is the discount factor; and $\phi$ are the parameters of the value function.

The expected return from state $s_t$ is given by Eq.\ref{eq_Gt}.
.\begin{equation}
G_t = R_t + \gamma \cdot V(s_{t+1};\phi) + \gamma^2 \cdot V(s_{t+2};\phi) + ...
\label{eq_Gt} %Eq.\ref{eq_}
\end{equation}

The policy objective is the negative logarithm of the policy, weighted by the advantage function, and is given by Eq.\ref{eq_J}.
\begin{equation}
J(\theta) = -\frac{1}{T} \sum_{t=1}^{T} \log \pi_{\theta}(a_t|s_t) A_t
\label{eq_J} %Eq.\ref{eq_}
\end{equation}
where $\pi_{\theta}(a_t|s_t)$ is the policy network, and $A_t = G_t - V(s_t;\phi)$ is the advantage function.
The gradient of the policy objective concerning the parameters of the policy network is given by Eq.\ref{eq_nabla_J}.
\begin{equation}
\nabla_\theta J(\theta) = -\frac{1}{T} \sum_{t=1}^{T} \nabla_\theta \log \pi_{\theta}(a_t|s_t) A_t
\label{eq_nabla_J} %Eq.\ref{eq_}
\end{equation}

This gradient is used to update the parameters of the policy network using the Adam optimization algorithm, a variant of stochastic gradient descent. The value function is updated using the mean squared error between the observed returns and the estimated returns.

The policy network is updated iteratively until convergence, at which point the policy network has learned an optimal policy for mapping states to actions. The A3C algorithm has been shown to achieve state-of-the-art performance on a range of benchmark reinforcement learning problems.

\section{Experiment and Discussion}

\subsection{Experimental Environment}
\begin{figure}[ht]\centering 
\includegraphics[width=1.0\textwidth]{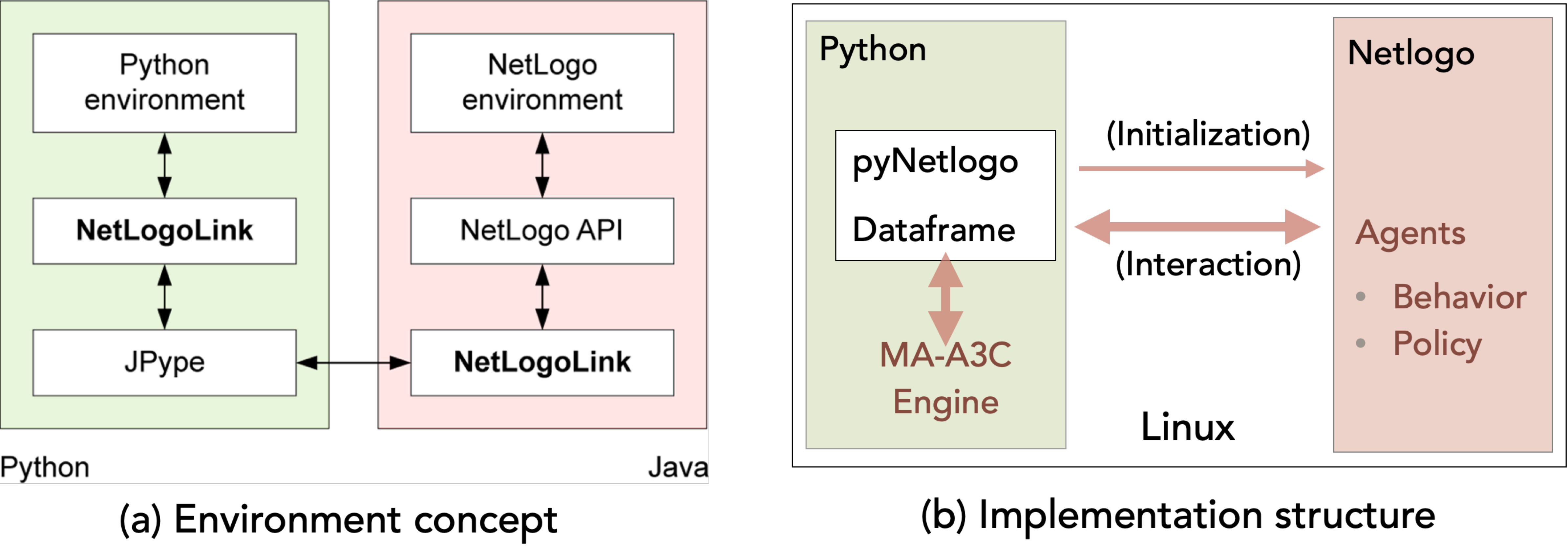}
\caption{Environment construction. (a) Environment concept shows the Multi-agent Reinforcement Learning (MARL) environment concept using Netlogo and Python. (b) Implementation structure shows the MA-A3C engine embedded system implementation structure in this study.}
\label{fig_sys_struc}
\end{figure}

\begin{figure}[ht]\centering 
\includegraphics[width=1.0\textwidth]{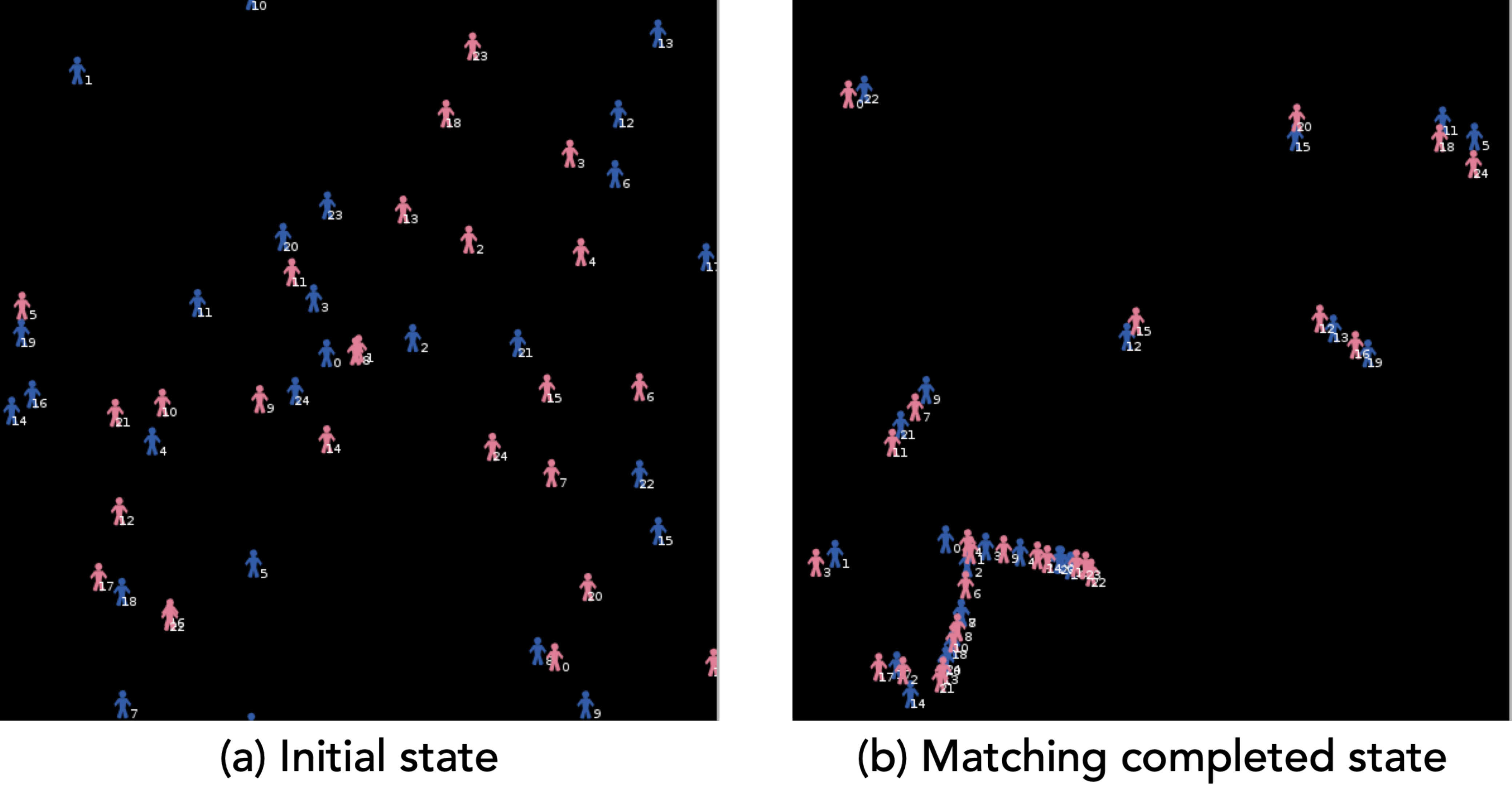}
\caption{Screenshots of the simulation are presented above, starting from the initial state and continuing until the end of the specified time step duration. The blue icons represent 'Male' agents, and the pink icons represent 'Female' agents. The agents act according to their own policies within a limited space.}
\label{fig_Screen}
\end{figure}

Assortative matching games are fundamental games that combine both competition and cooperation. In this study, we develop a simulation model, shown in Fig.\ref{fig_sys_struc}, using Netlogo (https://ccl.northwestern.edu/netlogo/) based on the "love coupling" idea introduced by Dan Ariely \cite{Ariely_upside_of_irrationality}. The model includes a set of basic strategies to simulate human behavior, with the simulation consisting of a population of male and female agents (default=25 each gender) divided into control and experimental groups. The control group only employs the basic strategies predefined in the model, while the experimental group can accept strategy optimization from the python-based MA-A3C reinforcement learning engine during the simulation, shown in Fig.\ref{fig_Screen}.

\subsection{Experiment Design}

\textbf{4.2.1 General Rules for Agents}\\
\\
Each agent has a unique numerical label (default value in the range [0,24]) and moves within a predefined activity space, searching for agents with higher numerical labels within their visual range. Agents can request pairing with others who accept or reject based on individual strategies. If pairing is successful, the reward value is equal to the sum of the numerical labels of both agents, which is shared equally by the successful agents. The aim of all agents is to obtain higher rewards.
\\\\
\textbf{4.2.2 Evaluation Criteria}\\
\\
The control group serves as a baseline for the study, representing the simulation pairing results without external intervention. The experimental group evaluates whether the MA-A3C reinforcement learning can achieve better-matching results, including whether the agents in the experimental group can successfully match with opposite-sex partners with higher numerical labels compared to those in the control group with the same numerical label. The agents in the experimental group serve as MA-A3C reinforcement learning workers, with their policy network adopting few steps update method and their global network adopting a round-based update method. This study explores the effectiveness of reinforcement learning algorithms in improving assortative matching game outcomes and provides valuable insights into the potential applications of reinforcement learning in social simulation.
\\\\
\textbf{4.2.3 Initial Policies and Learnable Variables}\\
\\
For all agents, the following initial strategies are adopted.

\paragraph{\textbf{Move policy}}
The agent selects a target. The target is the opposite-sex agent with the highest numerical label that has not yet successfully paired within the agent's visual range and whose label is not the most recent one in the 'being declined list' (to avoid the excessive pursuit of the same potential partner). If there are no eligible targets, the agent moves randomly. 

\paragraph{\textbf{Offer policy}}
The agent sends a pairing offer if the target is within a distance of three steps. At this point, both the sender and the recipient of the offer pause their movement. The recipient of the offer decides to accept or decline based on their decline and accept decision policy. 

\paragraph{\textbf{Decline policy}}
For the first $n$ offers received, the agent declines them unconditionally and records the sender's label in their own 'decline list'. The sender of the declined offer records the label of the target in their 'being declined list'. For offers received after the first $n$, if the sender's label is lower than the maximum value in the 'decline list', the offer is declined. 

\paragraph{\textbf{Accept policy}}
For the first $n$ offers received, the agent does not accept any of them. For offers received after the first $n$, if the sender's label is greater than or equal to the maximum value in the 'decline list', the offer is accepted, and the matching is completed.

\paragraph{\textbf{Policy optimization}}
Different experimental conditions were applied to the control and experimental groups during the game. The control group remained with the same initial policies and underwent no stochastic changes. In contrast, the experimental group underwent policy optimization, with their learnable variables, such as move direction, first decline $n$, decline list, and being declined list, updated accordingly.

\subsection{Experiment Result}

\subsubsection{Simulation Result of Control Group (Baseline)}
Based on the evaluation criteria described in Section 4.2.2, the experiments simulated the results of all agents in the control group. The goal was to analyze how \textbf{equal matching} was affected by different conditions, especially the view range of the agents. As shown in Table.\ref{tab:baseline}, a larger view range led to more equal matching. In other words, if some agents in the experimental group were able to match with partners whose numerical labels had a higher-than-average gap, it indicated that the MA-A3C algorithm was effective.

\begin{table}   % \cellcolor{yellow}  % \multicolumn{2}{c}
\centering
\caption{Baseline results of assortative matching on different view ranges}
\label{tab:baseline}
\begin{tabular}{ccp{12mm}p{12mm}p{12mm}p{12mm}p{12mm}ccccccc}
\hline
&View&\multicolumn{3}{c}\textbf{Trial}&& &Ttrials  \\
\text{Abs. label diff.} &range&\text{\#1} &\text{\#2} & \text{\#3} & \text{\#4} & \text{\#5} & \text{Avg.} \\
\hline
Avg. label diff.&5& 8.80 & 8.40 & 7.56 & 7.44 & 6.36 & 7.71 \\
Standard deviation && 5.94 & 5.88 & 5.93 & 6.19 & 4.70 & 5.73 \\
\hline

Avg. label diff.&15& 7.28 & 3.84 & 2.96 & 2.76 & 5.36 & 4.44 \\
Standard deviation && 6.41 & 3.31 & 3.22 & 3.41 & 3.76 & 4.02 \\
\hline

Avg. label diff.&25& 0.96 & 0.88 & 0.64 & 0.84 & 1.76 & 1.02 \\
Standard deviation && 1.14 & 1.01 & 0.95 & 1.28 & 1.56 & 1.19 \\
\hline
Note: & \multicolumn{7}{l}{\text{Abs. label diff.: Absolute label difference of paired agents.}} \\
 & \multicolumn{7}{l}{\text{Avg. label diff.: Average of Abs. label diff.}} \\
\end{tabular}

%Note:\text{Abs. label diff.: Absolute label difference of paired agents}
%\text{Avg. label diff.: Average of Abs. label diff.}

\end{table}

\subsubsection{MA-A3C Reinforcement Learning for Policy Optimization}

The results of the baseline experiment showed that equal matching was relatively better with a view range of 25, making it appropriate to conduct a mixed experiment with a control and experimental group. We introduced an experimental group (optimized by the MA-A3C reinforcement learning) to examine how much assortative matching changed under this condition. We used the same 50x50 space and agent initialization, similar to the baseline experiment. The only difference was that we selected the smallest 10 labels (i.e., labels in [0, 9]) of male and female agents as the experimental group, which was tested within the same time step duration. After the matching was completed, compared to the baseline, the absolute difference of each pairing increased from 1.02 to 7.09, and the variance from 1.19 to 5.80. This indicates that the experimental group agents with smaller label values increased their competitiveness after policy optimization and could match with partners with larger label values. Table.\ref{tab:MA-A3C} shows the detailed results of the experiment conducted five times.

\begin{table}   % \cellcolor{yellow}  % \multicolumn{2}
\centering
\caption{Simulation results of the control group and experimental group in five trials (view range = 25).}
\label{tab:MA-A3C}
\begin{tabular}{cp{10mm}p{9mm} p{10mm}p{9mm}p{10mm}p{9mm}p{10mm}p{9mm}p{10mm}p{9mm}p{8mm}cc  cc cc cc cc}
\hline
&\multicolumn{2}{c}{\text{Trial \#1}} & \multicolumn{2}{c}{\text{\#2}}&\multicolumn{2}{c}{\text{\#3}} & \multicolumn{2}{c}{\text{\#4}}& \multicolumn{2}{c}{\text{\#5}}  \\
\hline
Rough steps&Female & Male & Female & Male &Female & Male & Female & Male & Female & Male \\
\hline
10&19 & \cellcolor{yellow}5 &22 & 18 &22 & 14 & 16 &\cellcolor{yellow} 7& 14 & 19 \\
&24  &22  &13  &12   &23  &13   &18   &19  &19   &17   \\
&20&14&18&24&24&22&19&24&12&21\\
&16&23&10&21&15&16&14&21&23&\cellcolor{yellow}4\\
100&21&24&20&20&21&\cellcolor{yellow}3&15&23&16&12\\
&\cellcolor{yellow}7&20&15&\cellcolor{yellow}5&19&\cellcolor{yellow}1&22&15&11&15\\
&17&21&23&\cellcolor{yellow}6&17&\cellcolor{yellow}2&13&18&12&\cellcolor{yellow}1\\
&\cellcolor{yellow}1&17&16&16&13&15&\cellcolor{yellow}5&20&15&\cellcolor{yellow}6\\
&\cellcolor{yellow}2&16&11&11&18&24&20&11&18&18\\
200&\cellcolor{yellow}5&15&19&17&16&12&23&17&22&\cellcolor{yellow}9\\
&18&\cellcolor{yellow}8&17&21&14&\cellcolor{yellow}7&\cellcolor{yellow}4&20&24&13\\
&15&\cellcolor{yellow}7&24&\cellcolor{yellow}6&11&17&24&\cellcolor{yellow}1&17&14\\
&22&\cellcolor{yellow}5&\cellcolor{yellow}1&14&\cellcolor{yellow}2&20&11&12&21&\cellcolor{yellow}5\\
&23&\cellcolor{yellow}2&21&10&\cellcolor{yellow}1&11&\cellcolor{yellow}3&16&\cellcolor{yellow}8&24\\
300&13&11&\cellcolor{yellow}8&15&20&10&21&\cellcolor{yellow}5&20&\cellcolor{yellow}3\\
&11&15&\cellcolor{yellow}4&13&12&\cellcolor{yellow}8&\cellcolor{yellow}2&9&\cellcolor{yellow}2&10\\
&0&2&7&8&4&21&12&10&10&8\\
&14&10&14&9&10&\cellcolor{yellow}9&10&8&\cellcolor{yellow}9&20\\
&8&\cellcolor{yellow}3&9&7&\cellcolor{yellow}9&19&\cellcolor{yellow}8&\cellcolor{yellow}4&7&7\\
400&10&9&12&\cellcolor{yellow}5&6&5&7&3&\cellcolor{yellow}5&11\\
&9&6&6&\cellcolor{yellow}1&5&4&9&6&4&2\\
&4&1&3&4&\cellcolor{yellow}8&23&17&\cellcolor{yellow}7&\cellcolor{yellow}0&16\\
&6&4&2&1&\cellcolor{yellow}7&18&0&2&6&5\\
&12&\cellcolor{yellow}0&5&2&3&6&6&3&3&0\\
500&3&3&0&0&0&0&1&0&1&2\\
\hline
C.Gr.Avg.Diff.&-2.63&-1.90&-1.80&-1.07&-2.73&-2.07&-1.90&-1.57&-3.00&-1.80\\
E.Gr.Avg.Diff.&2.10&4.70&1.00&3.30&4.10&3.10&1.90&3.30&2.60&4.60\\
\hline

Avg. label diff.&\multicolumn{2}{c}{\text{7.32}} & \multicolumn{2}{c}{\text{5.52}}&\multicolumn{2}{c}{\text{7.92}} & \multicolumn{2}{c}{\text{7.16}}& \multicolumn{2}{c}{\text{7.52}} \\
STD deviation&\multicolumn{2}{c}{\text{5.84}} & \multicolumn{2}{c}{\text{5.32}}&\multicolumn{2}{c}{\text{6.13}} & \multicolumn{2}{c}{\text{5.69}}& \multicolumn{2}{c}{\text{6.02}} \\

\hline
Note: &\multicolumn{10}{l}{\text{C.Gr.Avg.Diff.: Control group average difference of paired agents.}} \\
 &\multicolumn{10}{l}{\text{E.Gr.Avg.Diff.: Experimental group average difference of paired agents.}} \\
 & \cellcolor{yellow}&\multicolumn{9}{l}{\text{represents the outstanding agent in experimental group,}} \\
  & &\multicolumn{9}{l}{\text{which gains a higher label partner than group average.}} \\

\end{tabular}

\end{table}

\subsection{Discussion}

In our experiments, we found that view range had the most significant effect on the absolute deviation of matched pairs, as shown in Table.\ref{tab:baseline}. In a non-sparse space, a larger view range typically means more information can be obtained. This information represents potential competitive advantages for agent actions. Individuals with wider fields of view have more reference points for their behavior when they obtain more external information, enabling them to leverage their potential competitive advantages. This characteristic is also reflected in macrosocial and economic behavior, where increased information flow is conducive to better resource flow and matching.

Although the control group and experimental group had the same view range and were in the same environment, the agents in the experimental group had more competitive advantages through the MA-A3C optimization strategy. Specifically, the agents in the experimental group are the "workers" in the policy algorithm. The global network selectively learns their actions and policy experiences in the local environment. This selection is achieved through the multi-head attention mechanism. The experimental results show that not all agents in the experimental group performed well every time. Due to many objective random factors in the environment, a good action result is likely due to luck rather than strategy. Therefore, although the multi-head attention mechanism cannot completely filter and exclude this random luck, it can effectively give more attention to factors less affected by randomness during learning.

From the prior knowledge of the Secretary problem\cite{Secretary_problem} in human behavior analysis, we know that the optimal cutoff tends to $n/e$ as n increases, and the best applicant is selected with probability $1/e$. The MA-A3C engine can also obtain similar experiences in repeated reinforcement learning, thereby improving the agent's initial setting of $n$.

From the records of the experimental group's target selection, we found that sometimes agents follow the same-sex agents with high label values instead of the high label values of the opposite-sex agents in the conventional strategy. Opposite-sex agents favor same-sex agents with high label values, and following them may have more opportunities to interact with high-label value opposite-sex agents. Similar wisdom can also be found in human behavior patterns. Education can be understood as students following teachers to a certain extent.

\section{Conclusion}

In most environments handled by reinforcement learning, such as Atari2600, the agent's and environment's relationship is independent. However, in human behavior modeling, the association is more complex and richer.

This paper chose a fundamental human behavior pattern in which the agent's actions entirely determine the environment. The results of this modeling imply similar phenomena of human behavior in the real world. For example, under the economic man hypothesis, follow-up and collaborative strategies are the dynamic basis for forming an entrepreneurship and management team. In this study, we propose the MA-A3C algorithm, which combines the A3C algorithm with the multi-head attention mechanism to optimize the policy of a small number of agents in the experimental group and break the balance of equal matching.

From the experiment, we realized that equal matching is likely a specific manifestation of the Nash equilibrium in human behavior patterns. However, the underlying game mechanism requires further in-depth study in future research. Learning to act in multi-agent systems has received attention primarily from game theory, focusing on algorithms that converge to the Nash equilibrium. Reinforcement learning, on the other hand, focuses on acting optimally in stochastic scenarios. The learning process of systems of intelligent interacting agents is highly complex due to the complexity of single-agent learning combined with their communications and networked information dynamics, which is a topic for future work.


\begin{thebibliography}{8}

\bibitem{Ariely_upside_of_irrationality}
Dan Ariely. The Upside of Irrationality: The Unexpected Benefits of Defying Logic at Work and at Home.
Harper, 2010

\bibitem{DeepMind_2013}
Vlad Mnih et al. Playing Atari with Deep Reinforcement Learning.
Proc. NIPS Deep Learning Workshop, 2013.
%https://arxiv.org/abs/1312.5602

\bibitem{DeepMind_2015}
Vlad Mnih et al. Human-level control through deep reinforcement learning.
Nature 518, 529–533, 2015.

\bibitem{DeepMind_2016_A3C}
Vlad Mnih et al. Asynchronous Methods for Deep Reinforcement Learning.
Proc. International Conference on Learning Representations (ICLR), 2016

\bibitem{TRPO}
John Schulman et al. Proceedings of the 32nd International Conference on Machine Learning.
PMLR 37:1889-1897, 2015.

\bibitem{PPO}
John Schulman et al. Proximal Policy Optimization Algorithms. 2017

\bibitem{Deep_Energy_Based_Policies}
Matt Hoffman et al. Reinforcement Learning with Deep Energy-Based Policies. 
Proc. 34th International Conference on Machine Learning (ICML 2017)

\bibitem{SAC}
Tuomas Haarnoja et al. Soft Actor-Critic: Off-Policy Maximum Entropy Deep Reinforcement Learning with a Stochastic Actor. 
Proc. Conference on Neural Information Processing Systems (NeurIPS 2018)

\bibitem{ACUTE)}
Chengzhang Zhao et al.  Attentional Communication Framework for Multi-Agent Reinforcement Learning in Partially Communicable Scenarios.
Electronics 2022. 
%https://doi.org/10.3390/electronics11244204

\bibitem{MARL_Game}
L. Buşoniu et al. A comprehensive survey of multi-agent reinforcement learning.
IEEE Transactions on Systems, Man, and Cybernetics, Part C (Applications and Reviews), vol. 38, no. 2, pp. 156-172, Mar. 2008.

\bibitem{GT-RL)}
Kaveh Madani et al. A game theory–reinforcement learning (GT–RL) method to develop optimal operation policies for multi-operator reservoir systems.
Journal of Hydrology, Volume 519, Part A, 2014. %Pages 732-742, ISSN 0022-1694, https://doi.org/10.1016/j.jhydrol.2014.07.061.

\bibitem{GT-RL)}
Nowé, A. et al. Game Theory and Multi-agent Reinforcement Learning. 
Wiering, M., van Otterlo, M. (eds) Reinforcement Learning. Adaptation, Learning, and Optimization, vol 12. Springer. 2012
%https://doi-org.waseda.idm.oclc.org/10.1007/978-3-642-27645-3_14

\bibitem{Secretary_problem}
T. S. Ferguson. Who solved the secretary problem? Statistical Science, volume 4, pp.282-296. 1988.


\end{thebibliography}
%

\end{document}